\documentclass[prc,twocolumn,showpacs,10pt,floatfix]{revtex4}

\usepackage{graphicx}
\usepackage{amsmath}
\usepackage{amssymb}
\usepackage{xspace}

\newcommand{\timesSunits}{\ensuremath{\times10^{42}\,\mathrm{MeV}^{-2}\mathrm{s}^{-2}}\xspace}

\newcommand{\lt}{\ensuremath{<}}
\newcommand{\gt}{\ensuremath{>}}

\newcommand{\identity}[1]{#1}  

\begin{document}

\title{Simplified approach to the application of the geometric collective model}

\author{M. A. Caprio}
\affiliation{Wright Nuclear Structure Laboratory, Yale University, 
New Haven, Connecticut 06520-8124}

\date{\today}

\begin{abstract}
The predictions of the geometric collective model (GCM) for different
sets of Hamiltonian parameter values are related by analytic scaling
relations.  For the quartic truncated form of the GCM~--- which
describes harmonic oscillator, rotor, deformed $\gamma$-soft, and
intermediate transitional structures~--- these relations are applied
to reduce the effective number of model parameters from four to two.
Analytic estimates of the dependence of the model predictions upon
these parameters are derived.  Numerical predictions over the entire
parameter space are compactly summarized in two-dimensional contour
plots.  The results considerably simplify the application of the GCM,
allowing the parameters relevant to a given nucleus to be deduced
essentially by inspection.  A precomputed mesh of calculations
covering this parameter space and an associated computer code for
extracting observable values are made available through the Electronic
Physics Auxiliary Publication Service.  For illustration, the nucleus
$^{102}$Pd is considered.

\end{abstract}

\pacs{21.60.Ev, 21.10.Re, 27.60.+j}
\maketitle

\section{Introduction}

The geometric collective model (GCM) of Gneuss, Mosel, and
Greiner~\cite{gneuss1969:gcm,gneuss1970:gcm,gneuss1971:gcm,eisenberg1987:v1}
provides a description of nuclear quadrupole surface excitations using
a Schr\"odinger-like Hamiltonian.  The Hamiltonian is expressed as a
series expansion in terms of the surface deformation coordinates
$\alpha_{2\mu}$ and the conjugate momenta $\pi_{2\mu}$,
\begin{equation}
\begin{split}
\label{eqngcmhamiltonian}
H=&\frac{1}{B_2}[\pi\times\pi]^{(0)}+B_3[[\pi\times\alpha]^{(2)}\times\pi]^{(0)}+\cdots\\
&+C_2[\alpha\times\alpha]^{(0)}+C_3[[\alpha\times\alpha]^{(2)}\times\alpha]^{(0)}+\cdots.
\end{split}
\end{equation}
The terms may be classified as either ``kinetic energy'' terms,
involving the momenta $\pi$, or ``potential energy'' terms, involving
only the coordinates $\alpha$.  If only the lowest-order kinetic
energy term is retained, the eigenproblem for this Hamiltonian reduces
to the Schr\"odinger equation in five-dimensional space, and the
kinetic energy operator is equivalent~\cite{eisenberg1987:v1} to that
of the Bohr Hamiltonian~\cite{bohr1998:v2}.  The potential energy
depends only upon the shape of the nucleus, not its orientation in
space, and can thus be expressed purely in terms of the Bohr-Mottelson
shape coordinates $\beta$ and $\gamma$~\cite{bohr1998:v2}, as
\begin{multline}
\label{eqnpotlseries}
V(\beta,\gamma)= \frac{1}{\sqrt{5}} C_2\beta^2 - \sqrt{\frac{2}{35}}
C_3 \beta^3 \cos{3\gamma} +\frac{1}{5}C_4\beta^4\\
-\sqrt\frac{2}{175}C_5\beta^5\cos3\gamma+\frac{2}{35}C_6\beta^6\cos^2
3\gamma +\frac{1}{5\sqrt{5}}D_6\beta^6+\cdots.
\end{multline}
The eigenproblem is commonly solved by diagonalization in a basis of
harmonic oscillator wave
functions~\cite{hess1980:gcm-details-238u,troltenier1991:gcm}.

Once wave functions for the nuclear eigenstates are calculated,
electromagnetic matrix elements can be evaluated.  The most commonly
used expression for the electric quadrupole operator is deduced using
the assumption that the nuclear charge is uniformly distributed within
a radius
$R=R_0(1+\sum_\mu\alpha_{2\mu}Y^*_{2\mu})$~\cite{eisenberg1987:v1,hess1981:gcm-pt-os-w},
which leads to a series expression
\begin{equation}
\label{eqnqe2coll}
Q_{2\mu}=\frac{3ZR_0^2}{4\pi}\left[\alpha^*_{2\mu}
-\frac{10}{\sqrt{70\pi}}[\alpha\times\alpha]^{(2)\,*}_\mu+\cdots
\right],
\end{equation}
where $R_0$$\equiv$$r_0A^{1/3}$ (with $r_0$=1.1\,fm in
Ref.~\cite{hess1981:gcm-pt-os-w}).

Attempts have been made to derive the parameters in the collective
Hamiltonian operator~(\ref{eqngcmhamiltonian}) from models of the
underlying single particle dynamics (\textit{e.g.},
Refs.~\cite{mosel1968:gcm-microscopic,kumar1974:150sm152sm-ppq,eisenberg1976:v3}).
However, the necessary theory is not sufficiently well developed to
provide a full description of the nuclear phenomenology.  An
alternative, more pragmatic approach is to choose the collective
Hamiltonian so as to best reproduce observed nuclear properties.  The
GCM Hamiltonian with eight parameters ($B_2$, $B_3$, $C_2$, $C_3$,
$C_4$, $C_5$, $C_6$, and $D_6$) accomodated by the existing codes is
capable of describing a rich variety of nuclear structures and can
flexibly reproduce many details of potential energy surface
shapes~\cite{vonbernus1975:gcm,eisenberg1987:v1}.

Manual selection of parameter values in this full eight-parameter
model is impractical, so parameter values for the description of a
particular nucleus must be found through automated
fitting~\cite{hess1980:gcm-details-238u,troltenier1991:gcm} of the
nuclear observables.  This process introduces technical difficulties
associated with reliable minimization in eight-dimensional space, and
often the parameter values appropriate to a nucleus are
underdetermined by the available observables~\cite{eisenberg1987:v1}.
As discussed in Ref.~\cite{hess1980:gcm-details-238u}, the available
data are usually sufficient to establish the qualitative nature of the
GCM potential and determine the coefficients of lower order terms,
while the higher-order parameters produce only fine adjustments to the
predicted structure.

It is therefore often desirable to use a more tractable form of the
model, in which the GCM Hamiltonian is truncated to the leading-order
term in the kinetic energy and to the three lowest-order terms in the
potential,
\begin{multline}
\label{eqnhtrunc}
H=\frac{1}{B_2}[\pi\times\pi]^{(0)} \\ +\frac{1}{\sqrt{5}} C_2\beta^2
- \sqrt{\frac{2}{35}} C_3 \beta^3 \cos{3\gamma}
+\frac{1}{5}C_4\beta^4.
\end{multline}
This Hamiltonian is sufficient to produce rotor, oscillator, and
deformed $\gamma$-soft structures as well as various more exotic
possibilities involving shape coexistence (see
Refs.~\cite{eisenberg1987:v1,zhang1997:gcm-trunc}).  With fewer
parameters in the Hamiltonian, it is more feasible to survey the full
range of phenomena accessible in the parameter space, and the
parameter values applicable to a given nucleus are much more fully
determined by the available observables.  These benefits must be
weighed against the limitations inherent in using the truncated model:
the full generality of the GCM is forsaken, precluding, for instance,
the description of rigid triaxiality (\textit{e.g.},
Ref.~\cite{troltenier1996:gcm-ru}), and, even within its qualitative
domain of applicability, the truncated model can be expected to have
reduced flexibility in reproducing subtleties of the potential energy
surface.

The truncated form of the GCM Hamiltonian~(\ref{eqnhtrunc}) still
contains four parameters ($B_2$, $C_2$, $C_3$, and $C_4$).  The
relationship between a set of values for these parameters and the
structure of the resulting predictions is not evident without detailed
calculations.  It would be useful to have a model which covers the
full range of features needed for description of the physical system
but which simultaneously has a dependence upon its parameters which is
simple, qualitatively predictable by inspection, and directly
understandable.  It is therefore desirable to further simplify the GCM
parameter space, but
\textit{without} additional truncation of the model.

In the present work, analytic scaling relations are applied to reduce
the effective number of model parameters from four to two
(Section~\ref{secscaling}).  Analytic estimates of the dependence of
the model predictions upon these parameters are derived
(Section~\ref{secmapping}), and the model predictions over the entire
parameter space are compactly summarized in two-dimensional contour
plots (Section~\ref{secnumerical}).  The results presented
considerably simplify the application of the GCM, allowing the
parameters relevant to a given nucleus to be deduced essentially by
inspection.  A precomputed mesh of calculations covering this
parameter space and an associated computer code for extracting
observable values are made available through the Electronic Physics
Auxiliary Publication Service (EPAPS)~\cite{epaps}. For illustration,
the nucleus $^{102}$Pd is considered, incorporating recent
spectroscopic data (Section~\ref{secillustration}).

\section{Scaling properties}
\label{secscaling}

Two basic properties of the Schr\"odinger equation can considerably
simplify the use of the model.  These properties relate the GCM
predictions for different sets of parameter values.

First, overall multiplication of any Hamiltonian by a constant factor
results in multiplication of all eigenvalues by that factor and leaves
the eigenstates unchanged.  This transformation leaves unchanged all
\textit{ratios} of energies, as well as all observables which depend only
upon the wave functions.  Therefore, all calculations can be performed
for some reference value of $B_2$, varying only $C_2$, $C_3$, and
$C_4$, and any calculation with another value of $B_2$ would be
equivalent to one of these to within a rescaling of energies.  The
number of active parameters in the truncated GCM Hamiltonian is
effectively reduced from four to three.

Second, it is a well-known heuristic that ``deepening'' a potential
lowers the energies of levels confined within the potential, while
``narrowing'' a potential raises the level energies.  It is thus
reasonable that successively deepening and then squeezing a given
potential, if performed in the correct proportion, could have effects
which offset each other.  For the $n$-dimensional Schr\"odinger
equation, of which the GCM eigenproblem with harmonic kinetic energy
is a specific case, this holds exactly.  Consider the transformation
in which the potential is multiplied by a factor $a^2$ while also
dilated by a factor $1/a$, \textit{i.e.}, multiplying $\beta$ by $a$
in the argument to $V$,
\begin{equation}
\label{eqnscalingpotential}
V'(\beta,\gamma)=a^2V(a\beta,\gamma).
\end{equation}
The effect of this transformation is simply to multiply all
eigenvalues by a factor $a^2$ and radially dilate the wave functions
by $1/a$ (see Appendix).  Since level energies are multiplied by the
same scaling factor $a^2$ as the potential itself, they retain their
positions relative to the recognizable ``features'' of the potential,
such as barriers or inflection points.  This scaling property allows a
further reduction of the number of active parameters in the truncated
GCM Hamiltonian from three to two, as described in the remainder of
this section.

If the electric quadrupole transition operator~(\ref{eqnqe2coll}) is
truncated to its linear term, then all matrix elements of this
operator change by the same factor, $a^{-1}$, under wave function
dilation (see Apendix).  Thus, all $B(E2)$ values are multiplied by
$a^{-2}$, and $B(E2)$ \textit{ratios} are left unchanged.  Many GCM
studies have retained the second-order term~\cite{troltenier1991:gcm},
but inclusion of this or other higher-order terms destroys the simple
invariance of $B(E2)$ ratios, since the different terms
in~(\ref{eqnqe2coll}) scale by different powers of $a$ under dilation.
The second-order term usually provides only a relatively small
correction to the linear term, and the correct coefficient by which it
should be normalized is highly
uncertain~\cite{eisenberg1987:v1,hess1981:gcm-pt-os-w}.  Comparative
studies by Petkov, Dewald, and Andrejtscheff~\cite{petkov1995:gcm-ba}
have shown no clear benefit to its inclusion for the nuclei
considered.  In light of the simple scaling properties obtained by its
omission, calculations of $B(E2)$ strengths are carried out using a
linear electric quadrupole operator throughout the present work.

A scaling result equivalent to that just discussed has been used in
Refs.~\cite{vonbernus1971:gcm-scaling,habs1974:gcm-n50to82,troltenier1991:gcm}
to simplify the automated fitting of experimental data with the full
eight-parameter GCM Hamiltonian.  The canonical transformation
$\pi\rightarrow\frac{1}{a}\pi$ and $x\rightarrow a x$ was used to
produce wave function dilation with no change in energy scale,
equivalent to the transformation~(\ref{eqnscalingpotential}) followed
by an overall multiplication of the Hamiltonian by $a^{-2}$.  However,
in
Refs.~\cite{vonbernus1971:gcm-scaling,habs1974:gcm-n50to82,troltenier1991:gcm},
the second-order form of the quadrupole operator was used, so $B(E2)$
ratios \textit{were} changed under dilation.  Therefore, the fitting
procedure could only be carried out using energy ratios, and, after
fitting, the wave functions were dilated to reproduce a single $B(E2)$
strength or quadrupole moment.

Systematic use of the scaling relations just discussed is facilitated
by the adoption of a simple reparametrization of the truncated GCM
potential, as
\begin{multline}
\label{eqnpotldef}
V(\beta,\gamma)=f \Biggl[
\frac{9}{112} d \left(\frac{\beta}{e}\right)^2 \\
- \sqrt{\frac{2}{35}} \left(\frac{\beta}{e}\right)^3 \cos{3\gamma} +
\frac{1}{5} \left(\frac{\beta}{e}\right) ^4
\Biggr].
\end{multline}
This expression has been constructed so that varying each of the
parameters $d$, $e$, and $f$ controls one specific aspect of the
potential:
\begin{enumerate}
\item[$d$~--] \underline{d}etermines the shape of the potential, \textit{i.e.}, $d$
uniquely defines the shape to within scaling
\item[$e$~--] \underline{e}xpands the potential horizontally, \textit{i.e.},
varying $e$ scales $V$ in the radial coordinate $\beta$
\item[$f$~--] is a \underline{f}actor multiplying the entire potential, \textit{i.e.},
varying $f$ scales the magnitude of $V$.
\end{enumerate}
With this choice of parameters, the transformations of the
potential~--- dilation and overall multiplication~--- necessary for
the application of the scaling properties are achieved simply by
varying $e$ or $f$.  Comparison of the coefficients in
(\ref{eqnpotldef}) with those in the original parametrization
(\ref{eqnhtrunc})
yields the conversion formulae
\begin{equation}
\label{eqndefccc}
\begin{alignedat}{3}
d &= \frac{112}{9\sqrt{5}}\frac{C_2C_4}{C_3^2}&
\qquad e&=\frac{C_3}{C_4}&
\qquad f&=\frac{C_3^4}{C_4^3}\\
C_2&=\frac{9\sqrt{5}}{112}\frac{fd}{e^2}&
\qquad C_3&=\frac{f}{e^3}&
\qquad C_4&=\frac{f}{e^4}.
\end{alignedat}
\end{equation}

\begin{figure}
\begin{center}
\includegraphics*[width=1.0\hsize]{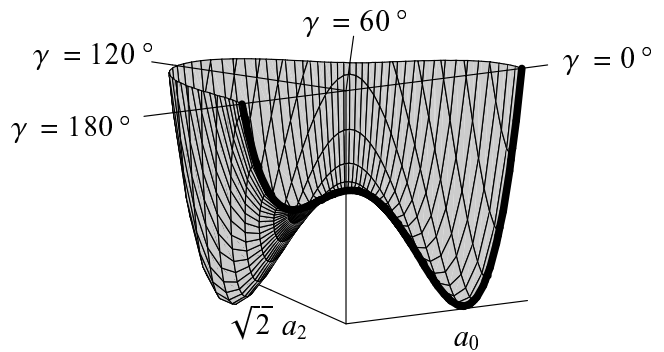}
\end{center}
\vspace{-6pt}
\caption
{The GCM potential function for $d$\lt0 posesses a global minimum, a
saddle point, and a local maximum at $\beta$=0.  The case $d$=$-5$ is
shown here as an example, plotted as a function of the Cartesian
coordinates $(a_0,\sqrt{2}a_2)$ or polar coordinates $(\beta,\gamma)$,
for $\gamma$=0$^\circ$ to 180$^\circ$.  The cut along the $a_0$-axis
(thick line) is the same as is shown in the second panel of
Fig.~\ref{figgcmpotld}.  The global minimum is visible in this figure
at $\gamma$=0$^\circ$ and 120$^\circ$, and the saddle point is visible
at $\gamma$=60$^\circ$ and 180$^\circ$.  The saddle point is a local
minimum with respect to $\beta$ and a local maximum with respect to
$\gamma$.
\label{figgcmsaddle}
}
\end{figure}

The extremum structure of the truncated GCM potential, investigated in
Refs.~\cite{greiner1963:gcm-dneg,acker1965:gcm-dpos}, can be expressed
very concisely in terms of the present parametrization
(\ref{eqnpotldef}).  Extrema occur where $V(\beta,\gamma)$ is locally
extremal with respect to both $\beta$ and $\gamma$ individually.
Thus, they are possible where $\cos 3\gamma$ attains its maximum value
of +1 (at $\gamma$=0$^\circ$, 120$^\circ$, and 240$^\circ$) or its
minimum value of $-1$ (at $\gamma$=60$^\circ$, 180$^\circ$, and
300$^\circ$).  Since the potential (\ref{eqnpotldef}) repeats every
120$^\circ$ in $\gamma$, it suffices to locate the extrema on one
particular ray in the $\beta\gamma$-plane with $\cos 3\gamma$=+1
(\textit{e.g.}, $\gamma$=0$^\circ$) and on one ray with $\cos
3\gamma$=$-1$ (\textit{e.g.}, $\gamma$=180$^\circ$).  Thus, as
illustrated in Fig.~\ref{figgcmsaddle}, extrema need only be sought on
a cut through the potential along the $a_0$-axis, and these extrema
will then be duplicated along the other rays of $\cos 3\gamma$=$\pm1$.
\begin{figure*}
\begin{center}
\includegraphics*[width=1.0\hsize]{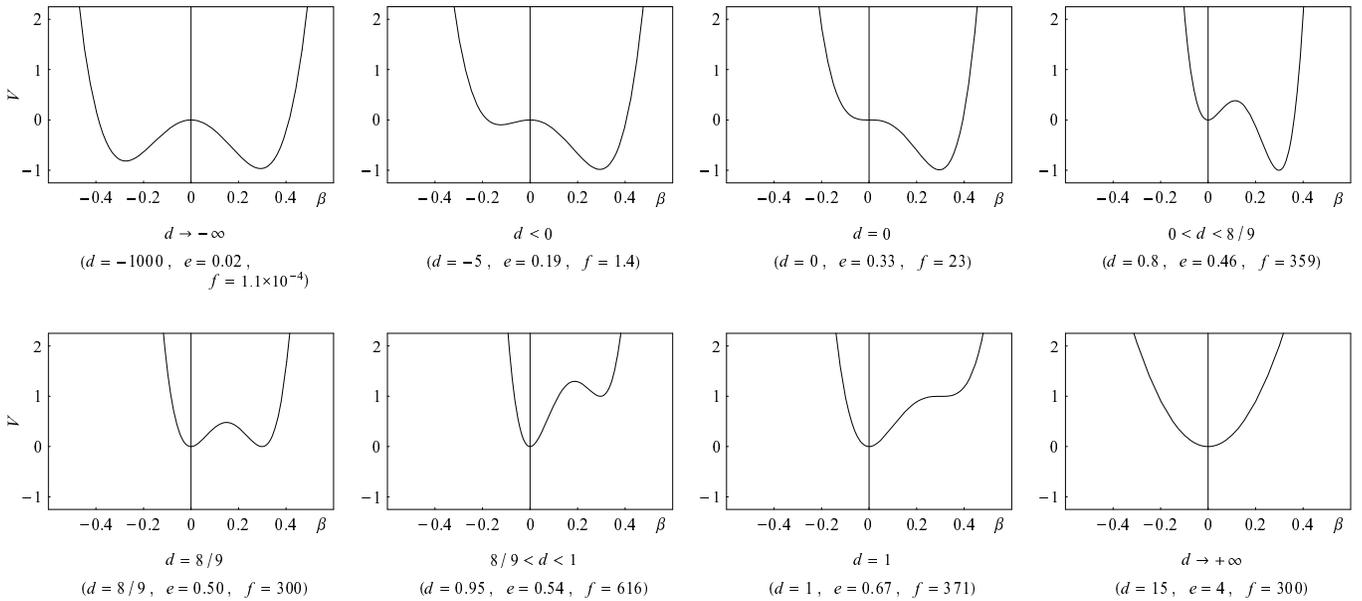}
\end{center}
\vspace{-12pt}
\caption{Illustration of the qualitatively different shapes of the GCM
potential function (\ref{eqnpotldef}) obtained for different ranges of
values for the parameter $d$.  Potentials are shown as a function of
$\beta$ along the $a_0$-axis cut (see text).  For $f$ in MeV, the
energy scale is also in MeV.
\label{figgcmpotld}
}
\end{figure*}
\begin{figure*}
\begin{center}
\includegraphics*[width=0.7\hsize]{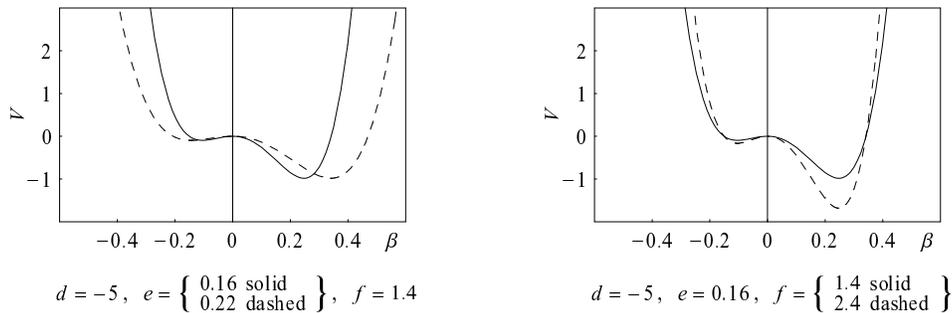}
\end{center}
\vspace{-12pt}
\caption
{Scalings of the GCM potential function (\ref{eqnpotldef}) obtained by
changing the values of the parameters $e$ and $f$.  The varying
parameter $e$ dilates the potential with respect to $\beta$, while
varying $f$ applies a multiplicative factor to the potential.
Potentials are shown as a function of $\beta$ along the $a_0$-axis cut
(see text).  For $f$ in MeV, the energy scale is also in MeV.
\label{figgcmpotlef}
}
\end{figure*}
\begin{figure}
\begin{center}
\includegraphics*[width=1.0\hsize]{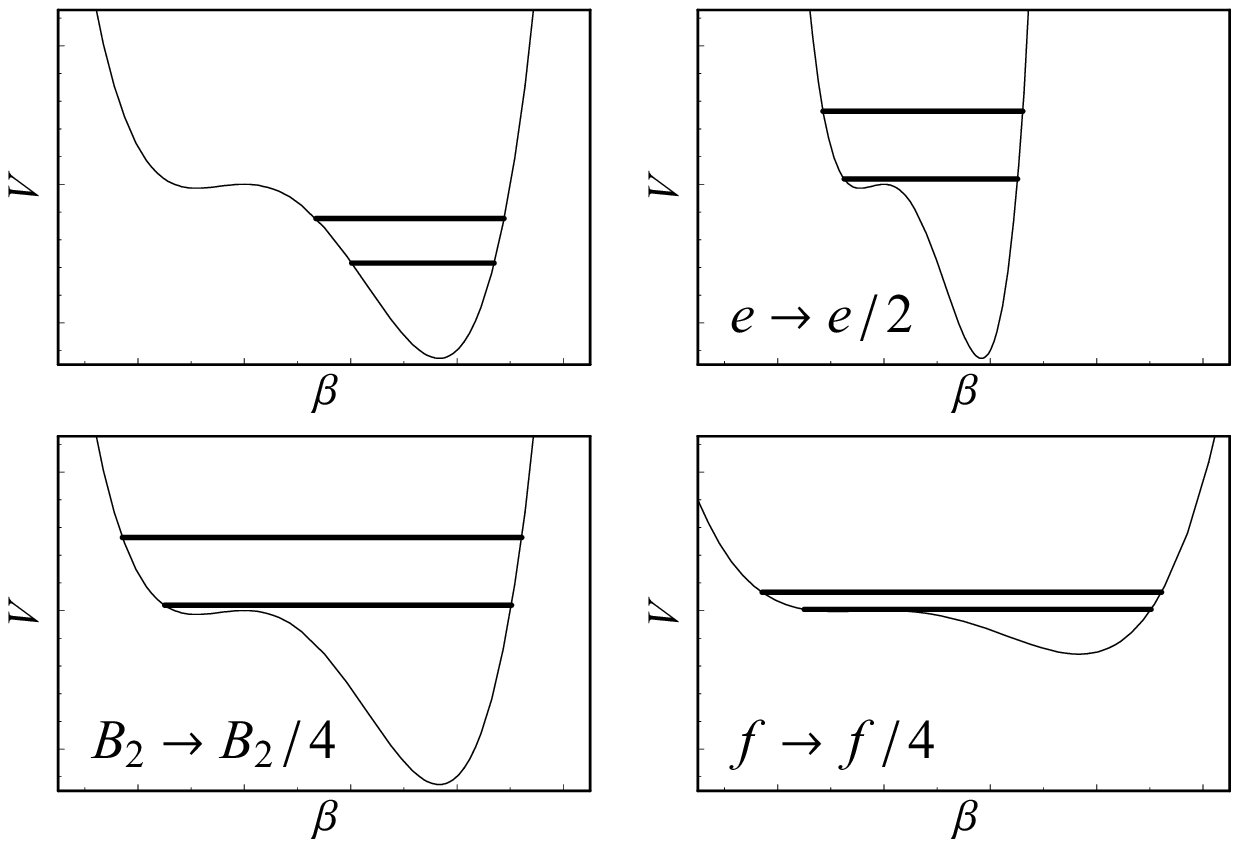}
\end{center}
\vspace{-6pt}
\caption
{Illustration of the effect of increasing $S$.  Increasing $S$ by a
factor of 4 relative to the original Hamiltonian (upper left) may be
accomplished in various ways: by decreasing the mass parameter $B_2$
by a factor of 4 (bottom left), by making the potential narrower by a
factor of 2 (top right), or by making the potential shallower by a
factor of 4 (bottom right).  The effects of each are equivalent (see
text).
\label{figseffects}
}
\end{figure}

Extrema along the two rays $\gamma$=0$^\circ$ and $\gamma$=180$^\circ$
are found by identification of the zeroes of $\partial
V/\partial\beta$ for $\cos 3\gamma$=+1 and for $\cos 3\gamma$=$-1$.
The variable $\beta$ is a radial coordinate and so takes on only
positive values.  However, the form of the following results is
considerably simplified by noting that the only occurrence of $\cos
3\gamma$ in (\ref{eqnpotldef}) is in a product also containing the
only occurrence of an odd power of $\beta$, so substituting $\cos
3\gamma$=$-1$ is algebraically equivalent to setting $\cos
3\gamma$=$+1$ and negating $\beta$.  Any extremum occuring for $\cos
3\gamma$=$-1$ will thus be found when the extrema for $\cos
3\gamma$=+1 are sought, but at a fictitious ``negative'' $\beta$
value.  A simple expression for the $\beta$ values yielding extrema
along $\gamma$=0$^\circ$ follows, but it must be interpreted with the
proviso that when a negative $\beta$ value is encountered it actually
represents a positive $\beta$ value along $\gamma$=180$^\circ$.  The
extrema of $V$ along the $a_0$-axis cut are located at
\begin{equation}
\beta=\begin{cases}
0, \beta_-, \beta_+ & d\leq 1 \\ 0 & d>1 \end{cases}
\end{equation}
where
\begin{equation}
\label{eqnbetapm}
\beta_\pm=\frac{3}{4}\sqrt{\frac{5}{14}}\left(1\pm r\right) e,
\end{equation}
in terms of $r$$\equiv$$\sqrt{1-d}$.  The extremal values of the
potential are
\begin{equation}
\label{eqnvbetapm}
V(\beta_\pm)=-\frac{135}{50176}(r\pm1)^3(3r\mp1)f.
\end{equation}

The nature of the extrema~--- whether they are minima, maxima, saddle
points, or inflection points~--- can be ascertained from the signs of
the partial derivatives.  The extremum structure of the potential
depends only upon the value of $d$, as summarized in
Fig.~\ref{figgcmpotld}.  For $d$\lt0, the potential has both a global
minimum and a saddle point at nonzero $\beta$
(Fig.~\ref{figgcmsaddle}).  For 0\lt$d$\lt1, minima are present at
both at nonzero $\beta$ and at $\beta$=0, with the deformed minimum
lower for 0\lt$d$\lt8/9 and the undeformed minimum lower for
8/9\lt$d$\lt1.  For $d$\gt1, there is only one minimum, located at
$\beta$=0.

Let us briefly address the ranges of definition for the parameters $e$
and $f$.  Negating $e$ reflects the potential about $\beta$=0.  A
positive value of $e$ places the deformed minimum on the prolate side
of the cut ($\beta_+$\gt0), while negative $e$ places the deformed
minimum on the oblate side of the cut ($\beta_+$\lt0).  All model
predictions for energies and transition strengths are unchanged under
interchange of prolate and oblate deformations, and only the signs of
quadrupole matrix elements and thus quadrupole moments are affected.
Throughout the discussions and examples in the present work, $e$ will
be taken positive without loss of generality.  Only positive values of
$f$ are meaningful, since for $f$ negative the coefficient on the
$\beta^4$ term in the potential is negative.  This makes
$V\rightarrow-\infty$ as $\beta\rightarrow\infty$, leaving the system
globally unbound.  The effects on the potential of varying the
parameters $e$ and $f$ are illustrated in Fig.~\ref{figgcmpotlef}.

In terms of the new parameters, overall multiplication of the
Hamiltonian by $b$ is obtained by the transformation
\begin{equation}
\label{eqntransfbdef}
B_2'=\frac{1}{b}B_2 \qquad d'=d \qquad e'=e \qquad f'=bf,
\end{equation}
and deepening the potential by $a^2$ while dilating by $1/a$ is
accomplished by the transformation
\begin{equation}
\label{eqntransfadef}
B_2'=B_2 \qquad d'=d \qquad e'=\frac{1}{a}e \qquad f'=a^2f.
\end{equation}
If two sets of parameter values, call them $(B_2,d,e,f)$ and
$(B_2',d',e',f')$, can be transformed into each other by any
combination of these scaling relations, the solutions for these
parameter sets will be identical, to within energy normalization and
wave function dilation.  If, however, $(B_2,d,e,f)$ and
$(B_2',d',e',f')$ cannot be transformed into each other
by~(\ref{eqntransfbdef}) and~(\ref{eqntransfadef}), the solutions for
these parameters will be distinct.  Parameter sets are thus naturally
grouped into ``families'', where $(B_2,d,e,f)$ and $(B_2',d',e',f')$
are members of the same family if and only if they are related by the
scaling transformations~(\ref{eqntransfbdef})
and~(\ref{eqntransfadef}).

We are now equipped to construct a ``structure parameter'' $S$ which
is invariant under the transformations~(\ref{eqntransfbdef})
and~(\ref{eqntransfadef}).  The quantity
\begin{equation}
\label{eqngcms}
S\equiv\frac{1}{B_2e^2f}
\end{equation}
may readily be verified to satisfy this condition.  If two points in
parameter space are characterized by the same values of $d$ and $S$,
they yield identical energy spectra, to within an overall
normalization factor, and identical wave functions, to within
dilation, and consequently identical $B(E2)$ ratios.  Two points
characterized by different values of $d$ or of $S$ will in general
give different energy spectra, wave functions, and $B(E2)$ ratios.

For a given potential shape, given by $d$, the parameter $S$
determines how ``high'' the levels lie relative to the features of the
potential.  An increase in $S$ may be achieved by decreasing the mass
parameter $B_2$, making the potential narrower, or making the
potential shallower.  Each of these has an equivalent effect on the
level spectrum~(Fig.~\ref{figseffects}), causing level energies to
rise relative to the potential.

\section{Mapping the GCM parameter space}
\label{secmapping}

The truncated GCM described in Section~\ref{secscaling} is effectively
a {\it two-parameter} model, with parameters $d$ and $S$.  The two
other degrees of freedom remaining from the four original parameters
provide only an overall normalization factor on the energy scale and,
through dilation of the wave functions in the coordinate $\beta$, on
the $B(E2)$ strength scale.  Because of the simplicity of this model,
the behavior of an observable over the {\it entire model space} can be
summarized on a single contour plot.  However, the parameter values
needed to cover the structural features of interest span many orders
of magnitude, so it is necessary to make some preliminary analytic
estimates to guide the numerical calculations if an effective and
comprehensive survey of the parameter space is to be made.
\begin{figure*}
\begin{center}
\includegraphics*[width=0.9\hsize]{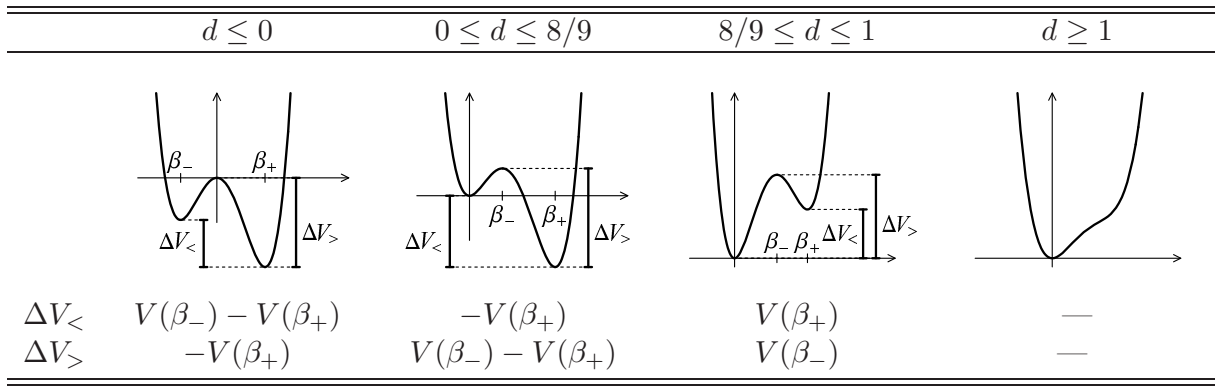}
\end{center}
\vspace{-12pt}
\caption
{Definitions of the energy differences $\Delta V_<$ and $\Delta V_>$
between extrema of the GCM potential, for different ranges of $d$
values, in terms of $V(\beta_\pm)$ [see~(\ref{eqnvbetapm})].
Different structure predictions are obtained for a given potential
shape depending upon the energy of the ground state and other
low-lying levels relative to the various extrema of the potential.
The quantities $\Delta V_<$ and $\Delta V_>$ are chosen so as to give
approximate ``boundaries'' on the level energies (taken relative to
the global minimum of the potential), separating the energies
corresponding to different structural regimes described in the text.
If low-lying levels have energies substantially less than $\Delta
V_<$, they are ``trapped'' in the global minimum.  If levels have
energies substantially greater than $\Delta V_>$, quartic oscillator
behavior dominates.  Potentials are plotted on the cut along the
$a_0$-axis.
\label{figdeltav}
}
\end{figure*}
\begin{figure}
\begin{center}
\includegraphics*[height=\hsize]{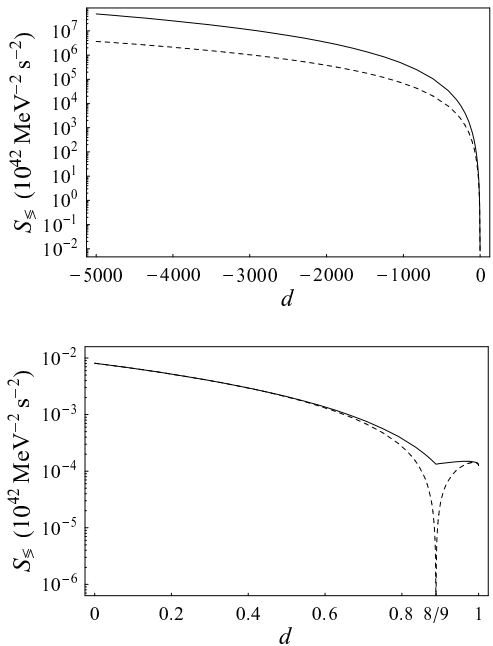}
\end{center}
\vspace{-18pt}
\caption
{Estimates $S_<(d)$ (dashed line) and $S_>(d)$ (solid line) of the $S$
values separating the different structural regimes (see text).  The
intervals $d\leq0$ and $0\leq d\leq1$ are plotted separately for
clarity.
\label{figslessgtr}
}
\end{figure}
\begin{figure*}
\begin{center}
\includegraphics*[width=0.8\hsize]{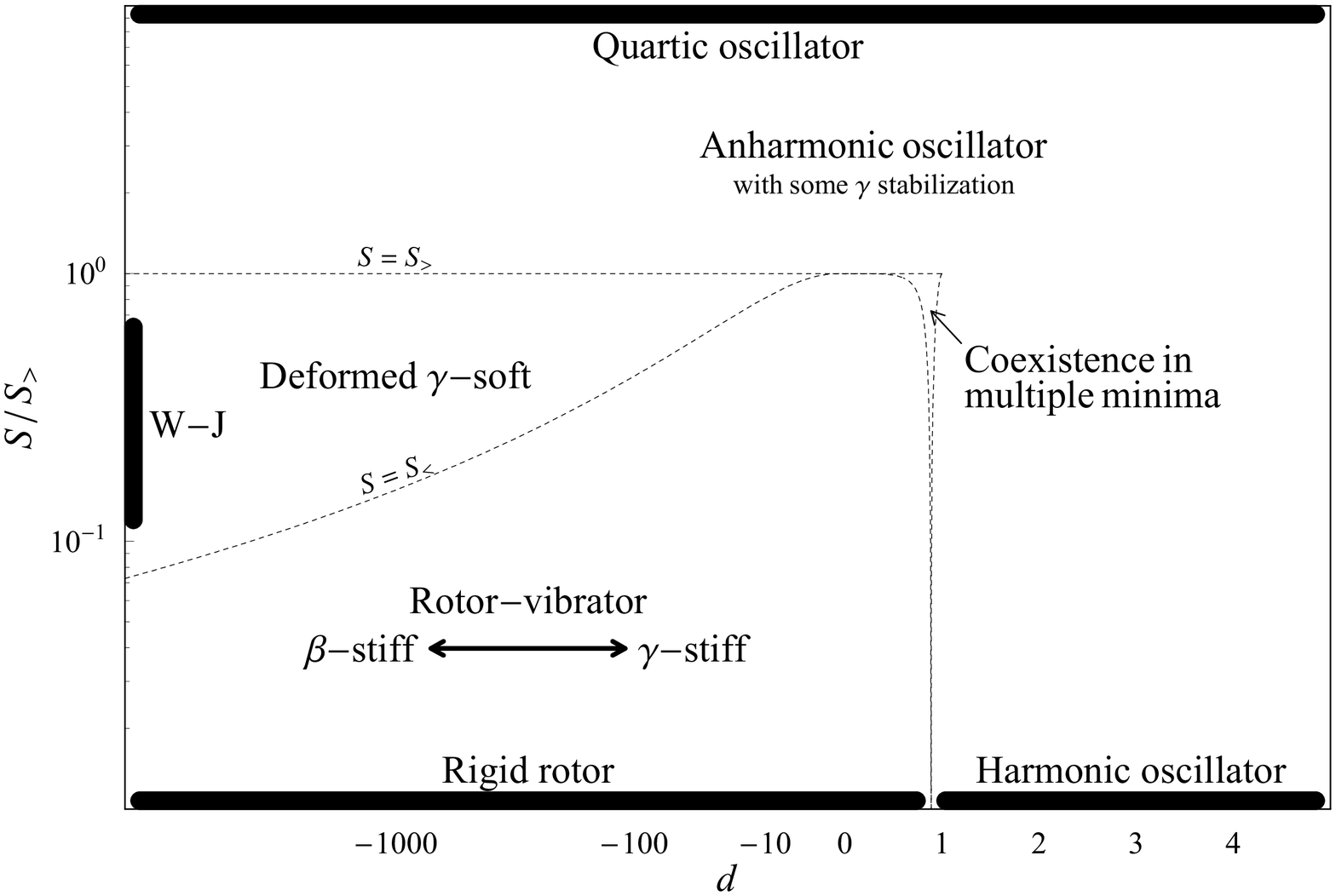}
\end{center}
\vspace{-12pt}
\caption
{Map of the GCM $(d,S)$ parameter space.  The regions in which
qualitatively different structures occur are indicated.  The curves
$S=S_<$ and $S=S_>$ (dotted lines) provide estimates for the
approximate boundaries between these regions.  Within the
rotor-vibrator region, the stiffness for $\beta$ and $\gamma$
vibrations varies with $d$ (double arrow) with a dependence given
approximately by (\ref{eqnbetagammaratio}).  Bars along the edges of
the plot represent structures which occur in their ideal form at
$d\rightarrow\pm\infty$ or at $S\rightarrow0$~or~$\infty$.  (``W-J''
denotes Wilets-Jean rigidly-deformed $\gamma$-soft
structure~\cite{wilets1956:oscillations}.)  The $d$- and $S$-axis
scales match those of the following figures
(Figs.~\ref{figgcmcontour_full_e} and~\ref{figgcmcontour_full_be2}) to
facilitate direct comparison with the calculated observable values.
\label{figgcmmap}
}
\end{figure*}

Let us first consider what qualitatively different types of behavior
are possible within the model space.  Each of the different potential
``shapes'' depicted in Fig.~\ref{figgcmpotld} can give rise to several
different types of structure, depending upon the excitation energy of
the ground state and other low-lying levels relative to the minimum of
the potential.  For potentials with $d$\lt0:
\begin{enumerate}
\item If level energies lie well below the saddle point, the states are
energetically confined to the deformed minimum, yielding rotational
behavior.
\item If level energies lie between the saddle point and the local
maximum at $\beta$=0, all $\gamma$ values are energetically
accessible, but $\beta$=0 is still not accessible.  In this case,
deformed $\gamma$-soft structure is possible.
\item If level energies lie well above the local
maximum at $\beta$=0, the potential controlling the behavior of these
states is dominantly a $\beta^4$ quartic oscillator well.
\end{enumerate}
For potentials with 0\lt$d$\lt1, two minima are present, one at zero
deformation and one at nonzero deformation:
\begin{enumerate}
\item If level energies lie well below the higher minimum,
the states are energetically confined to the global minimum, yielding
rotational behavior for 0\lt$d$\lt8/9 or approximately harmonic
oscillator behavior for 8/9\lt$d$\lt1.
\item If level energies lie above both minima but below the saddle point barrier, the
energetically-accessible regions around the two minima are separated
from each other by this barrier.  States involving mixing through the
barrier may be possible.
\item If level energies lie well above the 
barrier, the behavior again approaches that of a quartic oscillator.
\end{enumerate}
Finally, for potentials with $d$\gt1, deformed structure is not
possible.  If level energies are low in the well, so the states are
confined to a region of small $\beta$ where the $\beta^2$ term in the
potential dominates, harmonic oscillator behavior arises.  For level
energies much higher in the well, the $\beta^4$ term dominates,
yielding quartic oscillator behavior.

Thus, for the potentials with $d$\lt1, three different structural
regimes are possible at each given value of $d$, and the qualitative
nature of the low-lying levels is expected to depend upon the
excitation energy of these levels relative to specific extremum
features of the potential well.  Fig.~\ref{figdeltav} defines energy
differences $\Delta V_<$ and $\Delta V_>$ describing characteristic
energy scales for the structural regimes. If the ground state and
other low-lying states occur at energies (measured relative to the
lowest point of the potential) substantially below $\Delta V_<$, the
structure is expected to be that of the lowest-energy regime.  For
energies near or between $\Delta V_<$ and $\Delta V_>$, structure in
the intermediate regime is possible.  And, for energies well above
$\Delta V_>$, the structure is that of the highest-energy, or quartic
oscillator, regime.

Let us estimate the $S$ values, at a given value of $d$, which place
the low-lying level energies in each of these ranges.  The
Wentzel-Kramers-Brillouin approximation (\textit{e.g.},
Ref.~\cite{messiah1999:qm}) yields a quantization condition
\begin{equation}
\label{eqnfivedimwkb}
\int_0^{\beta_\text{max}}d\beta
\sqrt{B[E-V(\beta)]}\approx\left(n+\frac{3}{4}\right)\hbar\pi,
\end{equation}
$n=0,1,2,\ldots$, on the radial coordinate in the five-dimensional
Schr\"odinger equation, ignoring here the five-dimensional equivalent
of the centrifugal potential, where the mass $B$ appearing in the
usual form of the Schr\"odinger equation is $\sqrt{5}B_2/2$.  Since we
seek only an order-of-magnitude estimate, let us replace $E-V(\beta)$
in~(\ref{eqnfivedimwkb}) by a constant value $E-\overline{V}$,
representing the excitation energy relative to an ``average'' floor of
the well, and consider only the ground state ($n$=0).
Then~(\ref{eqnfivedimwkb}) reduces to the condition,
\begin{equation}
\beta_\text{max}\approx\frac{1}{2}\frac{2\pi\hbar}{\sqrt{2B(E-\overline{V})}},
\label{eqndebroglie}
\end{equation} 
or, simply, that approximately one-half of a de~Broglie wavelength
must fit within the width $\beta_\text{max}$ of the well.  The
appropriate ``width'' of the GCM potential for this estimate depends
upon how ``high'' the ground state lies within the well.  We are now
considering the boundaries of the structural regimes, for which the
ground state energy lies near the upper extrema in the potential.  The
location $\beta_+$ of the prolate minimum provides a reasonable
order-of-magnitude measure of the well width in this case (see
Fig.~\ref{figdeltav}).  Then (\ref{eqndebroglie}) indicates that $S$
values of approximately
\begin{equation}
\label{eqnsestimate}
S_\lessgtr\approx\frac{1}{\pi^2\hbar^2}\left(\frac{\Delta
V_\lessgtr}{f}\right)\left(\frac{\beta_+}{e}\right)^2,
\end{equation}
yield ground state energies at the borders $\Delta V_<$ and $\Delta
V_>$ of the structural regimes.  The values of $S_\lessgtr$ obtained
by substituting the expressions for $\Delta V_\lessgtr$ from
Fig.~\ref{figdeltav} are functions of $d$ only
(Fig.~\ref{figslessgtr}).

For $d$\lt8/9, rotational behavior occurs when the levels are at a
sufficiently low energy with respect to the potential
[$S$$\lesssim$$S_<(d)$].  For these rotational nuclei, additional
useful analytic estimates can be made of the quantitative dependence
of observables on $d$ and $S$.  For states sufficiently low-lying
(well-confined) in the deformed minimum, the structure approaches that
of small oscillations about a deformed equilibrium, which is described
analytically in the rotation-vibration model
(RVM)~\cite{faessler1965:rvm}.  This model provides simple
leading-order estimates for the rotational, $\beta$-vibrational, and
$\gamma$-vibrational energy scales.  The $2^+$ state energy for the
yrast band is determined by the moment of intertia, giving
\begin{equation}
E_{2g}\approx \frac{\hbar^2}{B \beta_0^2},
\end{equation}
where $\beta_0$ is the equilibrium deformation.  The
$\beta$-vibrational and $\gamma$-vibrational excitation energies are
determined by the curvature of the potential in each of these degrees
of freedom, yielding
\begin{equation}
\label{eqnrvmbetagamma}
E_\beta\approx\hbar\sqrt{\frac{V_{\beta\beta}}{B}}
\qquad
E_\gamma\approx\hbar\sqrt{\frac{V_{\gamma\gamma}}{B\beta_0^2}},
\end{equation}
where $V_{\beta\beta}$$\equiv$$\partial^2V/\partial\beta^2$ and
$V_{\gamma\gamma}$$\equiv$$\partial^2V/\partial\gamma^2$.  The
bandhead state energies are related to $E_\beta$ and $E_\gamma$ by
$E(0^+_\beta)\approx E_\beta$ and $E(2^+_\gamma)\approx
E_\gamma+\frac{1}{3}E_{2g}$~\cite{faessler1965:rvm}.

These estimates may be evaluated for the GCM potential
(\ref{eqnpotldef}) in a straightforward
fashion~\cite{caprio2003:diss}.  The vibrational energies normalized
to the yrast $2^+$ energy are
\begin{align}
\label{eqnbetagammas}
\frac{E_\beta}{E_{2g}}&\approx
\frac{135}{448\sqrt{7}}\sqrt{\frac{r(1+r)^5}{\hbar^2 S}}\\
\frac{E_\gamma}{E_{2g}}&\approx
\frac{135\sqrt{3}}{448\sqrt{7}}\sqrt{\frac{(1+r)^5}{\hbar^2S}},
\end{align}
in terms of $r$$\equiv$$\sqrt{1-d}$, and the ratio of the $\beta$ and
$\gamma$ vibration excitation energies is simply
\begin{equation}
\label{eqnbetagammaratio}
\frac{E_\beta}{E_\gamma}\approx\sqrt{\frac{r}{3}}.
\end{equation}
The ratios (\ref{eqnbetagammas})--(\ref{eqnbetagammaratio}) depend
only upon $d$ and $S$, as expected from the scaling properties of
Section~\ref{secscaling}.  These estimates provide guidance, needed
for the numerical calculations of the following section, as to both
the range of $d$ values of physical interest and the appropriate axis
scale or calculational mesh spacing for the parameter $d$.  Observe
that, by~(\ref{eqnbetagammaratio}), it is expected that
``$\beta$-stiff'' rotors, with $E_\beta$$>$$E_\gamma$, occur for
$d$\lt$-8$, while ``$\gamma$-stiff'' rotors, with
$E_\gamma$$>$$E_\beta$, occur for $d$\gt$-8$.

The combined results of this section give a detailed picture of the
qualitative characteristics expected for predictions of the truncated
GCM and provide quantitative estimates as to where in the $(d,S)$
parameter space these properties are to be found.  The results are
summarized graphically as a ``map'' of the parameter space in
Fig.~\ref{figgcmmap}.

\section{Numerical results}
\label{secnumerical}

\begin{figure*}
\begin{center}
\includegraphics*[width=1.0\hsize]{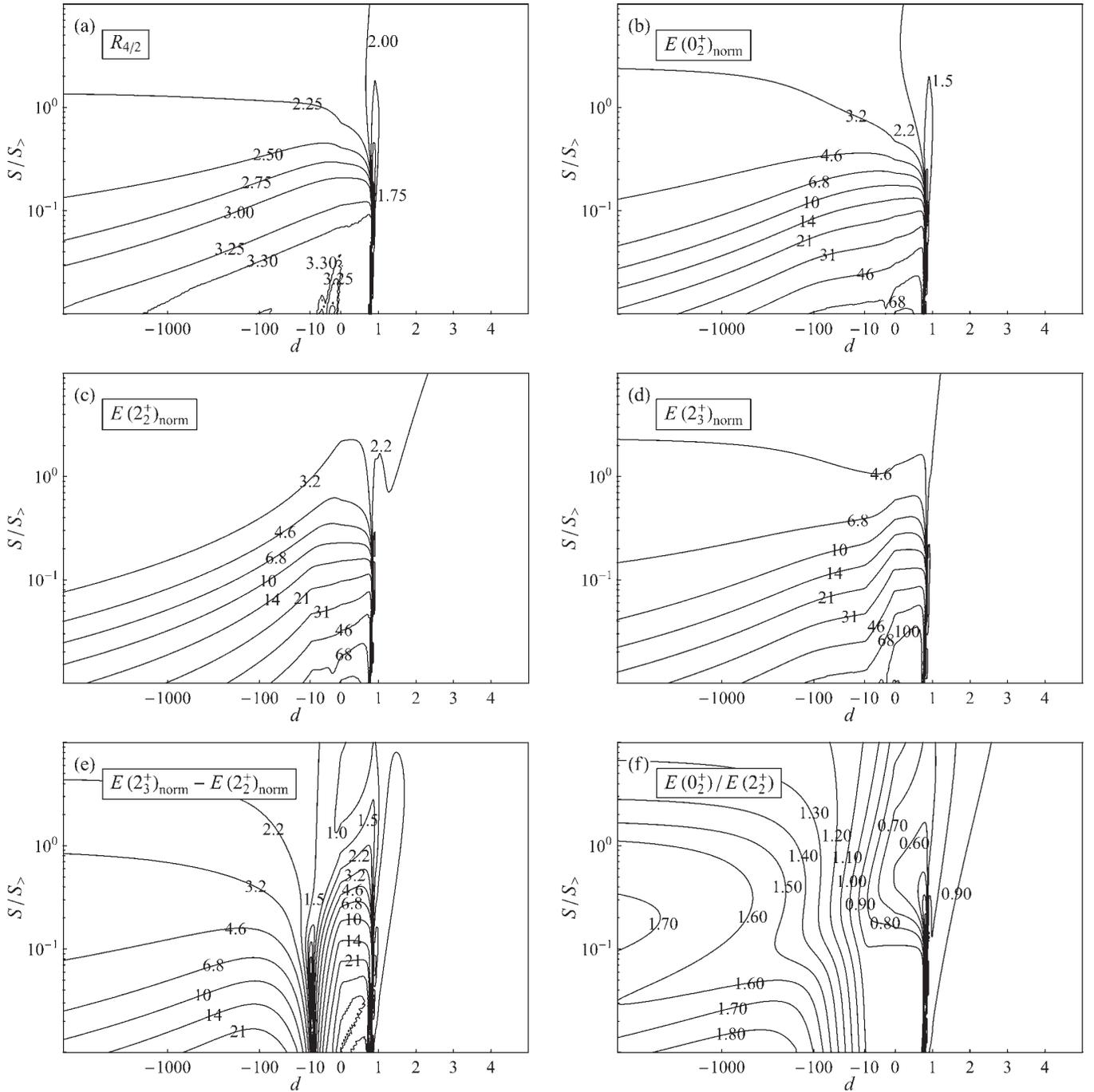}
\end{center}
\vspace{-12pt}
\caption
{Energy observable predictions of the GCM for $-5000\leq d \leq 5$ and
$10^{-2}\leq S/S_> \leq10^{+1}$:
(a)~$R_{4/2}$$\equiv$$E(4^+_1)/E(2^+_1)$,
(b)~$E(0^+_2)_\mathrm{norm}$$\equiv$$E(0^+_2)/E(2^+_1)$,
(c)~$E(2^+_2)_\mathrm{norm}$$\equiv$$E(2^+_2)/E(2^+_1)$, and
(d)~$E(2^+_3)_\mathrm{norm}$$\equiv$$E(2^+_3)/E(2^+_1)$.  For
examination of the crossing of the $\beta$ and $\gamma$ bands (see
text), it is also useful to plot the quantities
(e)~$[E(2^+_3)-E(2^+_2)]/E(2^+_1)$ and (f)~$E(0^+_2)/E(2^+_2)$.
\label{figgcmcontour_full_e}
}
\end{figure*}

\begin{figure*}
\begin{center}
\includegraphics*[width=1.0\hsize]{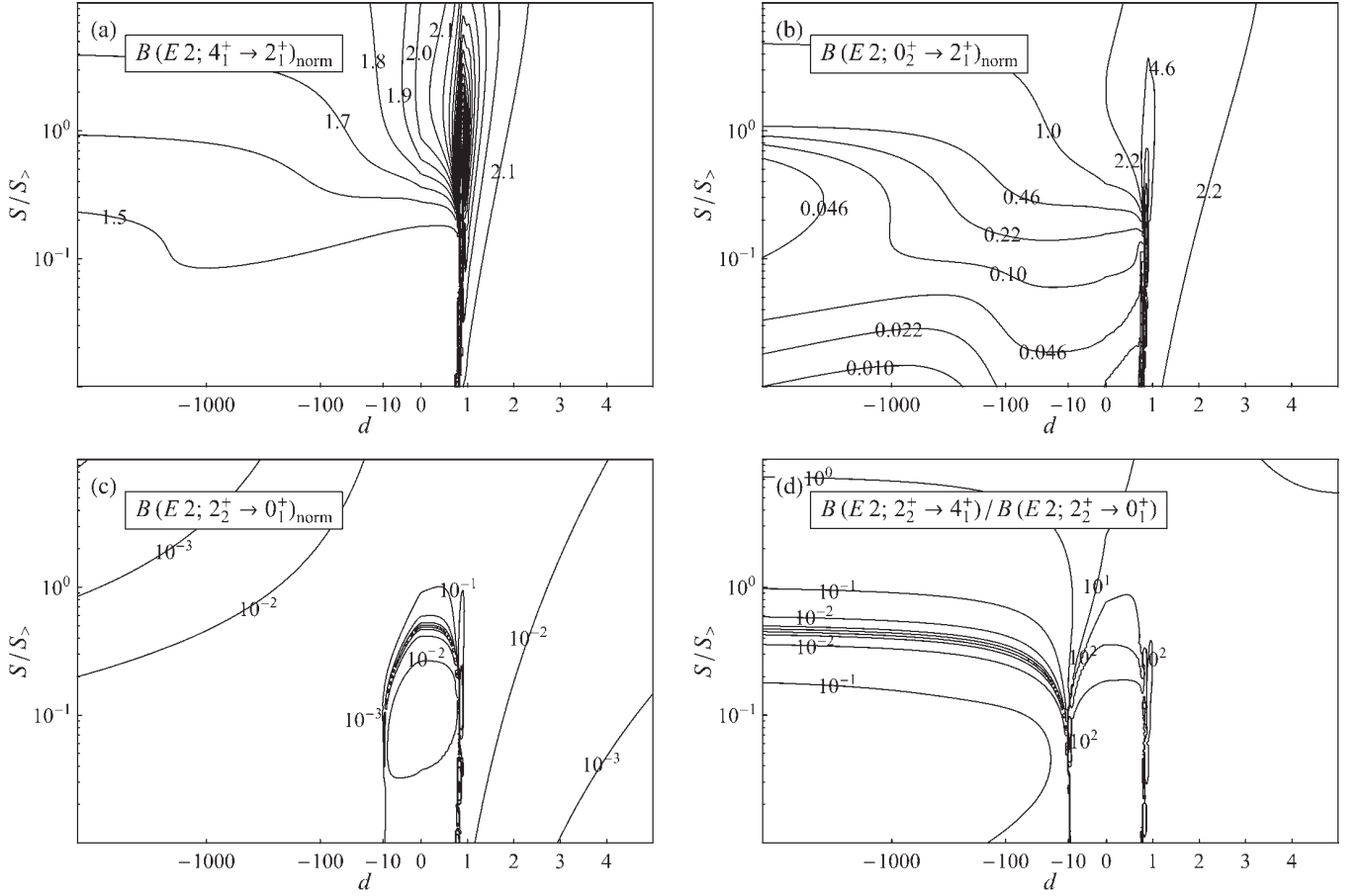}
\end{center}
\vspace{-12pt}
\caption
{$B(E2)$ observable predictions of the GCM for $-5000\leq d \leq 5$
and $10^{-2}\leq S/S_> \leq10^{+1}$:
(a)~$B(E2;4^+_1\rightarrow2^+_1)_\mathrm{norm}$$\equiv$$B(E2;4^+_1\rightarrow2^+_1)/B(E2;2^+_1\rightarrow0^+_1)$,
(b)~$B(E2;0^+_2\rightarrow2^+_1)_\mathrm{norm}$$\equiv$$B(E2;0^+_2\rightarrow2^+_1)/B(E2;2^+_1\rightarrow0^+_1)$,
(c)~$B(E2;2^+_2\rightarrow0^+_1)_\mathrm{norm}$$\equiv$$B(E2;2^+_2\rightarrow0^+_1)/B(E2;2^+_1\rightarrow0^+_1)$,
and (d)~$B(E2;2^+_2\rightarrow4^+_1)/B(E2;2^+_2\rightarrow0^+_1)$.
$B(E2)$ values are calculated using the linear form of the electric
quadrupole operator.
\label{figgcmcontour_full_be2}
}
\end{figure*}

Contour plots of several observables over the $(d,S)$ parameter space
are shown in Figs.~\ref{figgcmcontour_full_e}
and~\ref{figgcmcontour_full_be2}.  All observables plotted are
\textit{ratios} of energies or of $B(E2)$ values.  As discussed in the
previous sections, at a given $(d,S)$ any desired overall
normalization for the energy and $B(E2)$ scales can then be obtained
by proper rescaling of the well width, well depth, and mass parameter.
\begin{figure*}
\begin{center}
\includegraphics*[width=1.0\hsize]{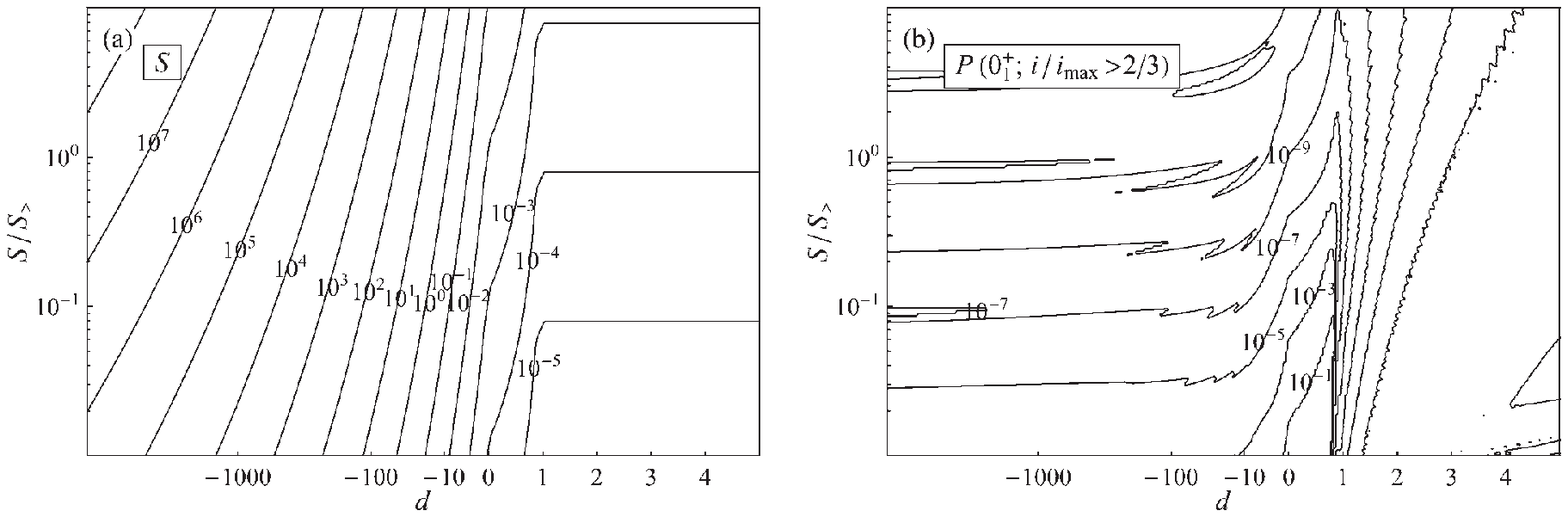}
\end{center}
\vspace{-12pt}
\caption
{Auxiliary plots for $-5000\leq d \leq 5$ and $10^{-2}\leq S/S_>
\leq10^{+1}$: (a)~$S$ values (in $10^{42}\,\mathrm{MeV}^{-2}\mathrm{s}^{-2}$) for the calculations of
Figs.~\ref{figgcmcontour_full_e} and~\ref{figgcmcontour_full_be2}, to
allow $S$ parameter values to be read directly, rather than as
$S/S_>$, for points on these plots, and (b)~the total probability
content of the highest (by phonon number) 1/3 of basis states for the
calculated $0^+_1$ state, as an indicator of convergence (see text).
\label{figgcmcontour_full_aux}
}
\end{figure*}

The $d$-axis in Figs.~\ref{figgcmcontour_full_e}
and~\ref{figgcmcontour_full_be2} extends from $d$=$-5000$ to 5.
Inclusion of the low end of this range is necessary to allow
description of rotational nuclei with high-lying $\beta$
vibrations~(\ref{eqnbetagammaratio}).  In order to encompass this full
range while maintaining a reasonably detailed view of the region
around $d$=0 in the plots, it is helpful to use a nonlinear $d$-axis
scale for $d\lt0$.  The estimate of $E_\beta/E_\gamma$
in~(\ref{eqnbetagammaratio}) indicates that this observable varies as
$(1-d)^{1/4}$, so for $d<0$ the $d$-axis of
Figs.~\ref{figgcmcontour_full_e} and~\ref{figgcmcontour_full_be2} is
chosen to be linear in $(1-d)^{1/4}$.

In the range of $d$ values being considered, the $S$ values resulting
in pheonomena of interest span approximately \textit{fourteen} orders
of magnitude, as seen from Fig.~\ref{figslessgtr}.  This occurs since
$S$ is defined in terms of $e$ and $f$, and the values of these
parameters needed to construct a reasonably-sized potential vary
greatly with $d$ (see Fig.~\ref{figgcmpotld} for examples).  However,
at any particular value of $d$, only about three decades in $S$, those
immediately surrounding $S$=$S_>(d)$, contain predictions of interest.
To make effective use of plotting space, the $S$-axis in
Figs.~\ref{figgcmcontour_full_e} and~\ref{figgcmcontour_full_be2} is
expanded to show only $10^{-2}S_>(d)\leq S
\leq10^{+1}S_>(d)$ at each point along the $d$-axis.
Fig.~\ref{figgcmcontour_full_aux}(a) facilitates the reading of $S$
directly off the contour plots.

Energy observables are shown in Fig.~\ref{figgcmcontour_full_e}.  The
values of the ratio $R_{4/2}\equiv
E(4^+_1)/E(2^+_1)$~[Fig.~\ref{figgcmcontour_full_e}(a)], an observable
which serves as a basic indicator of structure, closely match the
values expected from the $S_<$ and $S_>$ estimates.  The region with
2.2\lt$R_{4/2}$\lt2.6 corresponds approximately to that between the
dotted lines representing $S_<$ and $S_>$ in Fig.~\ref{figgcmmap}, in
which deformed $\gamma$-soft structure is expected.  For the
rotational-vibrational nuclei, found in the lower left-hand region of
the plots, the observables $E(0^+_2)/E(2^+_1)$, $E(2^+_2)/E(2^+_1)$,
and $E(2^+_3)/E(2^+_1)$ [Fig.~\ref{figgcmcontour_full_e}(b-d)] reflect
the basic dependences of the $\beta$ and $\gamma$ excitation energies
estimated in~(\ref{eqnbetagammas}) and~(\ref{eqnbetagammaratio}).  At
a given $d$, the excited band energies increase relative to $E(2^+_1)$
as $S$ decreases.  Degeneracy of the $\beta$ and $\gamma$ excitations
is expected at $d$$\approx$$-8$, and this behavior is clearly visible
from the sharp minimum in $[E(2^+_3)-E(2^+_2)]/E(2^+_1)$
[Fig.~\ref{figgcmcontour_full_e}(e)], where an avoided level crossing
occurs.  To the left of this division, the $2^+_2$ state is the
$\gamma$-vibrational bandhead; whereas, to the right of this division,
the $2^+_2$ state is the $\beta$-vibrational $2^+$ band member.
Proceeding to the left of the band crossing, the $0^+_2$ level energy
rises relative to the $2^+_2$ energy
[Fig.~\ref{figgcmcontour_full_e}(f)], as expected, due to increasing
$\beta$ stiffness.  The $0^+_2$ energy saturates at $\lesssim
2E(2^+_2)$, however, since as the $\beta$ vibrational excitation
continues to rise it leaves the two-phonon $\gamma$-vibrational state
as the lowest $K=0$ excitation (see Ref.~\cite{faessler1965:rvm}).

$B(E2)$ observable predictions are shown in
Fig.~\ref{figgcmcontour_full_be2}.  The ratio
$B(E2;4^+_1\rightarrow2^+_1)/B(E2;2^+_1\rightarrow0^+_1)$
[Fig.~\ref{figgcmcontour_full_be2}(a)] varies essentially smoothly
from rotational values to harmonic oscillator values, except that
extreme large and small values are encountered in a narrow region
between $d\approx0.8$ and $d\approx0.9$.  In this region, structures
involving coexistence in multiple minima are expected, so the $2^+_1$
and $4^+_1$ levels do not necessarily correspond to the same
structure.  Numerical convergence may also not be occurring for
certain calculations in this region, as discussed below.  The
observable $B(E2;0^+_2\rightarrow2^+_1)/B(E2;2^+_1\rightarrow0^+_1)$
[Fig.~\ref{figgcmcontour_full_be2}(b)] is of interest as the
$\beta$-vibrational decay strength for rotational nuclei.  In the
region $d$$\lesssim$$-8$,
$B(E2;2^+_2\rightarrow0^+_1)/B(E2;2^+_1\rightarrow0^+_1)$
[Fig.~\ref{figgcmcontour_full_be2}(c)] is the $\gamma$-vibrational
decay strength.  The predictions for the observable
$B(E2;2^+_2\rightarrow4^+_1)/B(E2;2^+_2\rightarrow0^+_1)$
[Fig.~\ref{figgcmcontour_full_be2}(d)] may be compared with the
``Alaga ratio'' predictions for true rotors~\cite{bohr1998:v2}, those
for the $2^+_\gamma$ state if $d$$\lesssim$$-8$ or for the $2^+_\beta$
state if $d$$\gtrsim$$-8$.

Numerical diagonalization of the GCM Hamiltonian is susceptible to
convergence
failure~\cite{gneuss1971:gcm,margetan1982:basis-optimization}.  The
code of Ref.~\cite{troltenier1991:gcm}, used for the present
calculations, obtains eigenvalues and eigenfunctions by
diagonalization in a basis of five-dimensional harmonic oscillator
wave functions.  If the set of basis functions chosen is not
sufficiently complete or is not adequately matched in radial extent to
the eigenfunctions it is being used to approximate, the process fails
to correctly calculate the GCM predictions for the eigenvalues and
eigenfunctions.  The basis is characterized by the phonon number $N$
at which it is truncated and by the stiffness parameter
$s\equiv(B_2C_2/\hbar^2)^{1/4}$ of the oscillator from which it is
constructed, which controls the radial extent of the basis functions.
For the present calculations, the basis was truncated at $N$=30, and
$s$ was automatically optimized by the variational procedures of
Refs.~\cite{margetan1982:basis-optimization,troltenier1991:gcm}.

Significant probability in the highest basis functions indicates that
the solution has ``overflowed'' the truncated basis and that the
results of the diagonalization are not reliable.  For successfully
convergent diagonalizations, most of the probability density is
concentrated in the ``lowest'' (by oscillator phonon number)
$\sim$10$\%$ of the basis
functions~\cite{gneuss1971:gcm,margetan1982:basis-optimization}.
Fig.~\ref{figgcmcontour_full_aux}(b) provides a contour plot of the
probability content of the ``highest'' 1/3 of basis states for the
calculated $0^+_1$ state.  Large values ($\gtrsim$$10^{-3}$) indicate
a calculation for which nonconvergence is likely to have occured.
However, this quantity is only a partial indicator of convergence, and
it does not provide a substitute for detailed convergence
studies~\cite{gneuss1971:gcm,margetan1982:basis-optimization} to
determine the actual extent of convergence for each excited state.

Convergence failure occurs for some of the most sharply deformed
structures covered by Figs.~\ref{figgcmcontour_full_e}
and~\ref{figgcmcontour_full_be2}.  At very low $S$ values in the rotor
region, where the most clearly well-deformed structure occurs and
$R_{4/2}$ should approach 3.33, the numerical calculations instead
exhibit sporadic patches of sharp fall-off in the $R_{4/2}$ value
[Fig.~\ref{figgcmcontour_full_e}(a), at $d\lesssim0$].  Comparison
with Fig.~\ref{figgcmcontour_full_aux}(b) shows that these
calculations exhibit relatively high probability contents for the
highest 1/3 of basis functions.  Convergence failure also occurs for
parameter sets near $d=0.8$, where coexistence in multiple minima
occurs.

Application of the results described in this article requires the use
of meshes of calculations to produce contour plots of observable
values, such as those of Figs.~\ref{figgcmcontour_full_e}
and~\ref{figgcmcontour_full_be2}.  To facilitate this process, a
database has been made available through the Electronic Physics
Auxiliary Publication Service (EPAPS)~\cite{epaps}, containing energy
and $B(E2)$ observables for low-lying states from a mesh of
calculations covering the ($d$,$S$) parameter space considered here.
An accompanying computer program is provided to extract observable
values from this database, calculate arbitrary expressions involving
these observables, and tabulate them in formats commonly accepted by
contour plotting codes.  The mesh consists of 150 steps along the
$d$-axis and 100 along the $S$-axis, uniformly distributed with
respect to the axis scalings described above.  Further details may be
found in the EPAPS documentation~\cite{epaps}.

\section{Illustration of the approach: $\mathbf{^{102}\mathrm{Pd}}$}
\label{secillustration}

The geometric collective model is most useful for the interpretation
of transitional nuclei, to which the simple models for limiting
structures (\textit{e.g.}, the harmonic oscillator or rotor models)
cannot readily be applied.  The Pd isotopes provide an example of a
spherical to deformed $\gamma$-soft structural transition and have
been extensively modeled as transitional nuclei using the interacting
boson model
(IBM)~\cite{stachel1982:ibm,bucurescu1986:ibm,pan1998:so6u5}, the
IBM-2~\cite{vanisacker1980:ibm,kim1996:ibm,giannatiempo1998:ibm}, and
other collective models~\cite{vorov1985:quartic}.
\begin{figure*}[t]
\begin{center}
\includegraphics*[width=0.9\hsize]{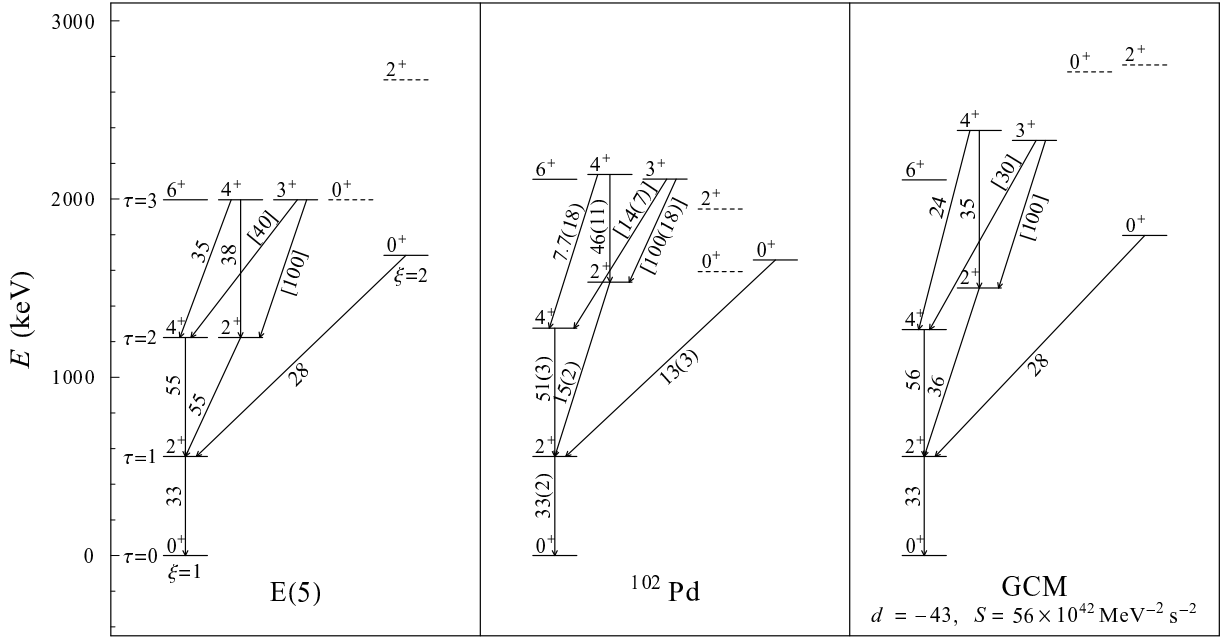}
\end{center}
\vspace{-12pt}
\caption{Experimental level scheme and
$B(E2)$ strengths for $^{102}$Pd, alongside the E(5) predictions and
the GCM predictions for $d$=$-43$ and $S$=56\timesSunits, normalized
to the experimental $E(2^+_1)$ and $B(E2;2^+_1\rightarrow0^+_1)$
values.  Observed levels with no clear theoretical counterpart or
calculated levels with no clear experimental counterpart (see text)
are indicated with dashed lines.  Experimental $B(E2)$ strengths are
from
Refs.~\cite{zamfir2002:102pd-beta,nds1998:102,luontama1986:102pd104pd-p2n-ppprime-coulex},
with the assumption of pure $E2$ multipolarity if the multipolarity is
not otherwise known.  $B(E2)$ values in brackets are relative values,
while all others are absolute values in W.u.
\label{fig102pdschemes}
}
\end{figure*}

Recent spectroscopic measurements on
$^{102}$Pd~\cite{zamfir2002:102pd-beta} have clarified the scheme of
low-lying levels and provided a detailed set of information on $B(E2)$
strengths among these levels.  The data of
Ref.~\cite{zamfir2002:102pd-beta} support the description of
$^{102}$Pd as $\gamma$-soft and softly-deformed with respect to
$\beta$.  The low-lying levels approximate an O(5) multiplet structure
(Fig.~\ref{fig102pdschemes}), as expected for a $\gamma$-independent
potential~\cite{rakavy1957:gsoft}.  A reasonable
description~\cite{zamfir2002:102pd-beta} is obtained with the E(5)
model of Iachello~\cite{iachello2000:e5}, in which the Bohr
Hamiltonian is used with a potential which is $\gamma$-soft and flat
(a square well) with respect to $\beta$.  There are, however,
discrepancies, which may be addressed using the present formulation of
the GCM.

Let us first summarize the description of $^{102}$Pd obtained in
Ref.~\cite{zamfir2002:102pd-beta}.  Levels occur at approximately the
correct energies to constitute the $4^+$-$2^+$ members of a $\tau$=2
multiplet and the $6^+$-$4^+$-$3^+$ members of a $\tau$=3 multiplet in
a $\gamma$-soft description (Fig.~\ref{fig102pdschemes}).  The
$R_{4/2}$ energy ratio value of 2.29 suggest a structure intermediate
between that of the harmonic oscillator ($R_{4/2}$=2.0) and a
Wilets-Jean rigidly-deformed $\gamma$-soft structure
($R_{4/2}$=2.5)~\cite{wilets1956:oscillations}. The data of
Ref.~\cite{zamfir2002:102pd-beta} indicate that the ``allowed''
$\Delta\tau$=$-1$ transitions from the $4^+_{\tau=3}$ and
$3^+_{\tau=3}$ levels are greatly enhanced over the ``forbidden''
$\Delta\tau$=$-2$ transitions, although some necessary $M1$/$E2$
mixing ratios have not been measured.

\begin{figure}
\begin{center}
\includegraphics*[width=\hsize]{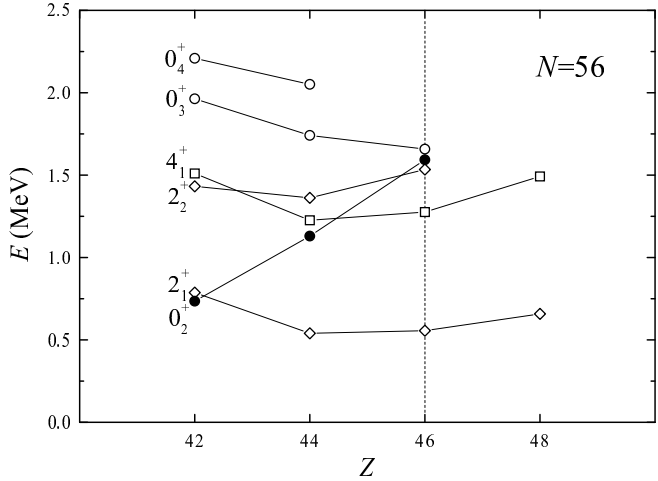}
\end{center}
\vspace{-12pt}
\caption{Evolution of level energies for low-lying states
along the $N$=56 isotonic chain.  The contrary behavior of the $0^+_2$
state (filled symbols), with a decreasing energy towards the $Z$=40
shell closure, indicates a cross-shell intruder nature.  The dotted
line indicates $Z$=46 (Pd).  (Figure based upon
Ref.~\cite{zamfir2002:102pd-beta}.)
\label{figsystn56}
}
\end{figure}

The low-lying, isomeric $0^+_2$ level at 1593\,keV is identified as a likely
intruder state.  An inspection of the evolution of level energies
along the $N$=56 isotonic chain reveals that the $0^+_2$ energy
\textit{decreases} towards the $Z$=40 shell closure
(Fig.~\ref{figsystn56}), indicating that this is not an ordinary
collective state constructed from the $Z$=40--50 valence space.  The
evolution is consistent with that of a collective excitation involving
the entire $Z$=28--50 shell, breaking the subshell closure at $Z$=40.
The next excited $0^+$ level, at 1658\,keV, lies at an energy 2.98
times the $2^+_1$ level energy, in near perfect agreement with the
E(5) prediction [$E(0^+_2)$=3.03$E(2^+_1)$].  This level decays with
collective $E2$ strength to the $2^+_1$ level
[13(3)\,W.u.], as expected for the head of
the $\xi$=2 family, though this strength is lower than predicted
[28(2)\,W.u.] by the E(5) model.  A $0^+$ member of the $\tau$=3
multiplet must be found for this picture to be complete.

However, several differences in quantitative detail are present
between experiment and the pure $\gamma$-soft predictions, as can be
seen from Fig.~\ref{fig102pdschemes}.  Some of the main deviations
from this picture are:
\begin{enumerate}
\item The $4^+$ and $2^+$
members of the proposed $\tau$=2 multiplet are split in energy, with
the $4^+$ level lower.
\item The absolute
$B(E2;2^+_{\tau=2}\rightarrow2^+_{\tau=1})$ strength is lower than the
$B(E2;4^+_{\tau=2}\rightarrow2^+_{\tau=1})$ strength.
\item Both the
$3^+_{\tau=3}$ and $4^+_{\tau=3}$ states show a strong preference to
decay to the $2^+_{\tau=2}$ state rather than to the $4^+_{\tau=2}$
state.  Only a slight such preference is expected for $\gamma$-soft
structure.
\end{enumerate}
These deviations from the E(5) model predictions are qualitatively
consistent with the incipient formation of a $\gamma$ band.  Such
structure would arise if the potential were perturbed from E(5) to
impose a slight preference for axial symmetry.

\begin{figure}
\begin{center}
\includegraphics*[width=\hsize]{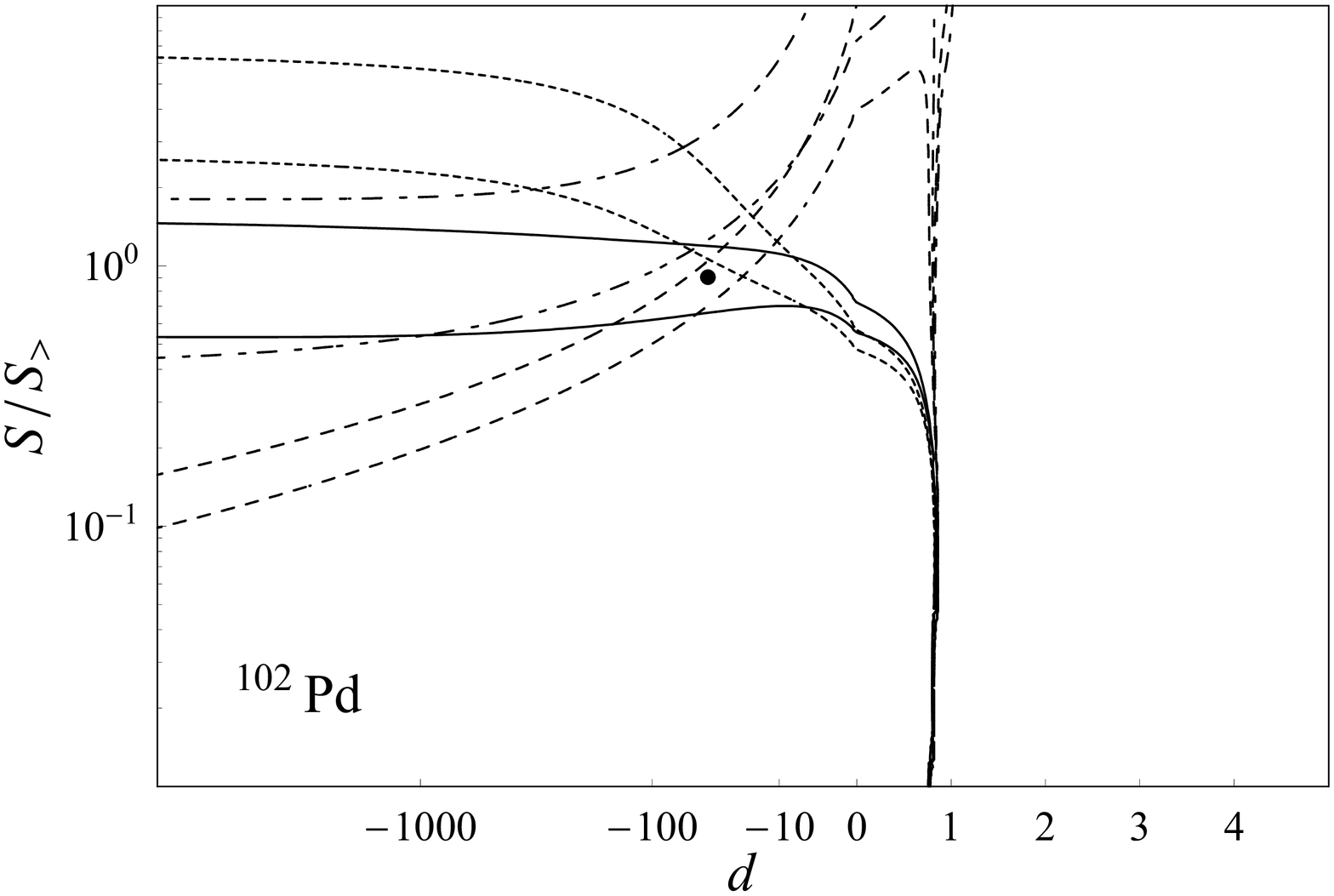}
\end{center}
\vspace{-12pt}
\caption{Regions in the ($d$,$S$) parameter space for which the
GCM predictions of selected observables match the values found in
$^{102}$Pd: $R_{4/2}$=2.29 to within 2$\%$ (solid line),
$E(2^+_2)/E(2^+_1)$=2.75 to within 5$\%$ (dashed line),
$E(4^+_2)/E(2^+_1)$=3.84 to within 5$\%$ (dashed-dotted line), and
$E(0^+_2)/E(2^+_1)$=2.98 to within 5$\%$ (dotted line).  The solid
circle indicates $d$=$-43$ and $S$=56\timesSunits, the parameter
values discussed in the text.
\label{fig102pdcontour}
}
\end{figure}

To consider this perturbation quantitatively, let us make use of the
truncated GCM.  The choice of values for the parameters $d$ and $S$ is
constrained simply by the requirement that the low-lying
energy observables be reproduced.  It is desirable to retain the good
agreement with the experimental $4^+_{\tau=2}$ level energy, $\tau$=3
multiplet energy, and first excited $0^+$ level energy obtained with
the E(5) description while also reproducing the splitting of the
$\tau$=2 multiplet.  Contours indicating the regions in GCM ($d$,$S$)
parameter space for which the predictions match the experimental
$R_{4/2}$, $E(2^+_2)/E(2^+_1)$, $E(4^+_2)/E(2^+_1)$, and
$E(0^+_2)/E(2^+_1)$ values are shown in Fig.~\ref{fig102pdcontour}.
We seek a point in parameter space simultaneously satisfying these
conditions.  Values for other energy and $B(E2)$ observables could be
used to refine the choice, but we are interested here primarily in the
general nature of the GCM predictions.  The parameter values
$d$$\approx$$-43$ and $S$$\approx$56\timesSunits constitute a
reasonable compromise.

\begin{figure}
\begin{center}
\includegraphics*[width=\hsize]{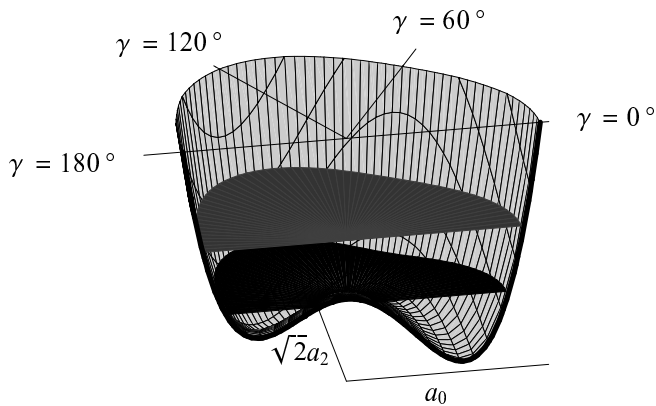}
\end{center}
\vspace{-12pt}
\caption{Plot of the potential energy surface
for the GCM calculation for $^{102}$Pd ($d$=$-43$, $S$=56\timesSunits)
as a function of the shape coordinates, showing the $0^+_1$ and
$0^+_2$ level energies (lower and upper laminae).
\label{fig102pdpotl}
}
\end{figure}

The GCM predictions for these values of $d$ and $S$ are shown in
Fig.~\ref{fig102pdschemes}~(right).  The $4^+_1$-$2^+_2$ splitting and
low-lying level energies are well reproduced, although some splitting
is introduced between the $6^+_1$, $4^+_2$, and $3^+_1$ ``multiplet''
levels.  The $B(E2)$ strengths for the $2^+_2\rightarrow2^+_1$,
$4^+_2\rightarrow4^+_1$, and $3^+_1\rightarrow4^+_1$ transitions are
all reduced relative to the E(5) predictions, while the other
transition strengths in Fig.~\ref{fig102pdschemes} are relatively
unaffected.  The change in the predicted strengths for these three
transitions is in the correct sense to bring them closer to the
experimental values but leaves these strengths still a factor of two
to three greater than observed.  As noted above, one of the other
outstanding issues regarding the interpretation of $^{102}$Pd in the
context of the E(5) picture is the nonobservation of any $0^+$ member
of the proposed $\tau$=3 multiplet.  In the GCM calculation, the
$0^+_3$ level is predicted to be substantially higher in energy than
the $6^+_1$, $4^+_2$, and $3^+_1$ states.
The potential for the GCM calculation for $^{102}$Pd is
plotted in Fig.~\ref{fig102pdpotl}, showing the extent
of the gentle minimum with respect to $\gamma$ at
$\gamma$=0$^\circ$.

\section{Conclusion}

The general approach and specific techniques discussed in this article
recast the GCM, with truncated Hamiltonian, as a very tractable model
for theoretical studies and practical application.  The use of
analytic scaling relations reduces the effective dimensionality of the
parameter space for this model, and basic estimates of the parameter
dependence of structural properties further simplify navigation of
this parameter space.  The parameter values most appropriate for the
description of a given nucleus can be deduced by inspection of contour
plots such as those in Figs.~\ref{figgcmcontour_full_e}
and~\ref{figgcmcontour_full_be2}.  To facilitate application of these
results, a database of calculations covering the model space and a
computer code for extracting observable values are provided through
the EPAPS~\cite{epaps}.

\vfill

\begin{acknowledgments}
Discussions with N.~V.~Zamfir, R.~F.~Casten, R.~Kr\"ucken,
E.~A.~McCutchan, and F.~Iachello are gratefully acknowledged.  This
work was supported by the US DOE under grant DE-FG02-91ER-40609.
\end{acknowledgments}

\appendix*
\section{}

An elementary proof is provided of the scaling property used in
Section~\ref{secscaling}, that, for the Schr\"odinger equation in $n$
dimensions, ``deepening'' and ``narrowing'' a potential in the correct
proportion simply multiplies all eigenvalues by an overall factor and
dilates all eigenfunctions.

Consider the Schr\"odinger equation in the $n$ Cartesian coordinates
$x_1,\ldots,x_n$, with a kinetic energy operator having the standard
Cartesian form and a potential energy operator which depends only upon
the coordinates,
\begin{multline}
\label{eqnse}
\left[\sum_{i=1}^n \left( - \frac{\hbar^2}{2 m_i} \frac{\partial^2}{\partial
x_i^2}\right) + V(x_1,\ldots,x_n) - E\right]\\
\times \Psi(x_1,\ldots,x_n)=0,
\end{multline}
where the $m_i$ are constants, $V$ is the potential energy operator,
and $E$ is the energy eigenvalue for wave function $\Psi$.  Suppose
that this equation is satisfied by a particular function $\Psi$, with
eigenvalue $E$, for a specific potential $V$.  Now consider the
related expression obtained by substituting the quantities
\begin{equation}
\begin{aligned}
\label{eqnprimedefs}
V'(x_1,\ldots,x_n)&=a^2V(ax_1,\ldots,ax_n)\\
\Psi'(x_1,\ldots,x_n)&=\sqrt{a^n}\Psi(ax_1,\ldots,ax_n)\\
E'&=a^2E
\end{aligned}
\end{equation}
for the corresponding quantities in (\ref{eqnse}).  This yields
\begin{multline}
\Bigg[\sum_{i=1}^n \Bigg( - \frac{\hbar^2}{2 m_i} \frac{\partial^2}{\partial
x_i^2}\Bigg) + a^2V(ax_1,\ldots,ax_n) - a^2E\Bigg]\\
\times\sqrt{a^n}\Psi(ax_1,\ldots,ax_n).
\end{multline}
With the substitution $u_i=ax_i$, this becomes
\begin{multline}
\Bigg[\sum_{i=1}^n \Bigg( - \frac{\hbar^2}{2 m_i} a^2\frac{\partial^2}{\partial
u_i^2}\Bigg) + a^2V(u_1,\ldots,u_n) - a^2E\Bigg]\\
\times \sqrt{a^n}\Psi(u_1,\ldots,u_n),
\end{multline}
which vanishes identically for all values of $(u_1,\ldots,u_n)$, by
(\ref{eqnse}).  Thus, the dilated wave function $\Psi'$ satisfies the
Schr\"odinger equation with the same kinetic energy operator as
in~(\ref{eqnse}) but with the modified potential $V'$, and does so
with eigenvalue $E'$.  The factor $\sqrt{a^n}$ included in the
definition of $\Psi'$ serves to preserve normalization with respect to
the volume element $dx_1\cdots dx_n$.

It is often convenient to use $n$-dimensional spherical coordinates
$(r,\Omega)$, where $r=(x_1^2+\cdots+x_n^2)^{1/2}$ and $\Omega$
represents the angular coordinates.  (These correspond to polar or
spherical coordinates in the specific cases $n$=2 or $n$=3,
respectively.)  In these coordinates, the transformation is
$V'(r,\Omega)=a^2V(ar,\Omega)$ and
$\Psi'(r,\Omega)=\sqrt{a^n}\Psi(ar,\Omega)$, which preserves
normalization with respect to the volume element $r^{n-1}drd\Omega$.

The matrix elements of the operator $r^q$ are encountered in the
determination of electromagnetic transition strengths, so it is useful
to derive their properties under dilation.  Consider the radial
integral
\begin{equation}
I^{AB}=\int_0^\infty r^{n-1} dr \, [\Psi^{A}(r)]^*\,r^q\,[\Psi^B(r)],
\end{equation}
where the angular variables have been suppressed for simplicity.  Then
the radial integral for the corresponding transformed eigenfunctions,
\begin{equation}
{I^{AB}}'=\int_0^\infty r^{n-1} dr \,
[\sqrt{a^n}\Psi^{A}(ar)]^*\,r^q\,[\sqrt{a^n}\Psi^B(ar)],
\end{equation}
is related to the original integral by
\begin{equation}
\label{eqniscaling}
{I^{AB}}'=a^{-q} I^{AB},
\end{equation}
as can be shown by means of a simple change of variable $u=ar$.  This
property applies to any operator homogeneous of order $q$ in the
coordinates, such as each term contributing to the quadrupole
operator~(\ref{eqnqe2coll}).

The leading-order kinetic energy term in the GCM
Hamiltonian~(\ref{eqngcmhamiltonian}), substituting the canonical
momenta $\pi_{2\mu}$$=$$-i\hbar\partial/\partial\alpha_{2\mu}$, is
$-\hbar^2/(\sqrt{5}B_2)\sum_\mu
\partial^2/(\partial\alpha_{2\mu}\partial\alpha^*_{2\mu})$.
This is not manifestly Cartesian, due to the presence of mixed partial
derivatives, but a simple change of
variables~\cite{rakavy1957:gsoft,eisenberg1987:v1} recasts it in
Cartesian form, so the above results may be used.  The ``radial''
coordinate to which the scaling property applies is $\beta$.

\vfill



\end{document}